\documentstyle{article}

 \renewcommand{\baselinestretch}{1}

 \newcommand{\zet}{{{\mbox{Z}}\!\!{\mbox{\bf Z}}}}
 \newcommand{\rr}{{{\mbox{I}}\!{\mbox{\bf R}}}}

\input{epsf}

\begin{document}
\bibliographystyle{unsrt}
\renewcommand{\baselinestretch}{2.0}
 
\parindent 0pt
 
\begin{center}
{\Large \bf 
 Root Lattices and Quasicrystals
}
\end{center}

\begin{center}
{\large 
 M. Baake}

{\small
 Institut f\"ur Theoretische Physik,
 Universit\"at T\"ubingen, Auf der Morgenstelle 14, 
 D-7400 T\"ubingen,W.-Germany,
 University of Tasmania, Dept. of Physics, GPO Box 252C, Hobart Tas 7001,
 Australia}

\vspace{5mm}

{\large 
 D. Joseph, P. Kramer, and M. Schlottmann}

{\small
 Institut f\"ur Theoretische Physik,
 Universit\"at T\"ubingen, Auf der Morgenstelle 14, 
 D-7400 T\"ubingen,W.-Germany}

\vspace{5mm}

{\small appeared: J.\ Phys.\ A: Math.\ Gen.\ {\bf 23} (1990) L1037-L1041}
 
\end{center}

\vspace{5mm}

\parindent15pt
\begin{center}
{\bf Abstract}
\end{center}
 
{\small
It is shown how root lattices and their reciprocals might serve as the
right pool for the construction of quasicrystalline structure models.
All non-periodic symmetries observed so far are covered in minimal
embedding with maximal symmetry.
}

\vspace{1cm}

\parindent15pt

For the construction of quasiperiodic tiling models by means of projection
from higher-dimensional periodic structures, the primitive hypercubic
lattices are the most frequently used ones. This has pragmatic reasons:
the lattice $\zet ^n$ is simple, possible for every integer $n$, and its
symmetry (point as well as space group) is well-known. However, the choice
of $\zet ^n$ is too restrictive because several patterns either require a
non-minimal embedding (like the Penrose pattern, Fig.\ 1a) or are even
impossible that way (like the triangle pattern, Fig.\ 1b). The question now is
how to select suitable candidates from the infinite pool of higher-dimensional
lattices which combine the minimal embedding of the crystallographically
forbidden symmetries with a systematic and simple description. 
It is well-known \cite{mermin} that there is only one Fourier module in the
plane for each noncrystallographic n-fold symmetry (up to n=46) and a triple
of Fourier modules in 3-space with icosahedral symmetry. Therefore, from the
viewpoint of general quasiperiodic densities, a single generating lattice
each is sufficient. However, from the tiling point of view, there are
several inequivalent examples per symmetry to be expected. Because a 
classification of tilings is not in sight, the question for a set of
fundamental examples arises again.

Following this path, one is almost automatically guided to hypercubic centerings
and to root lattices, or, as will turn out immediately, to root lattices and
their reciprocals. Hypercubic centerings (i.e., centerings of the primitive
hypercubic lattice with the full hypercubic point symmetry) do not exist
for $n=1$ and $n=2$: they give back the primitive case. For $n=3$ one has the
fcc and the bcc structure, the same being true for $n>4$, where they are called
F-type and I-type structure, respectively. Only $n=4$ shows a higher
symmetry: F-type and I-type are equivalent and they possess a 3-times
larger holohedry than the primitive lattice $\zet ^4$. This specific 
4-D lattice will be of some importance in what follows.

\begin{figure}
\centerline{ \epsfxsize=3.5in \epsfbox{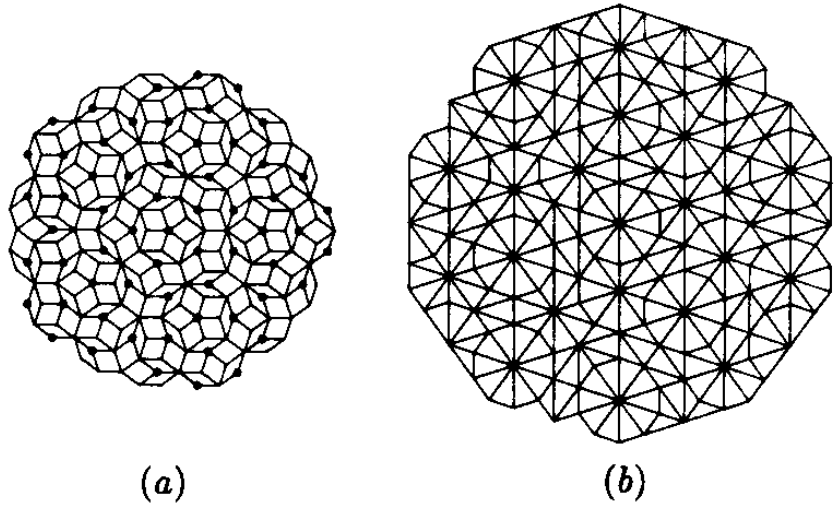}}
\caption{
 Quasiperiodic patterns as obtained from the root lattice
$A_4$, the Penrose pattern (a) with two classes of vertices and the triangle
pattern (b) with two different bond lengths.
}
\label{fig1}
\end{figure}

\begin{figure}
\centerline{ \epsfxsize=3.5in \epsfbox{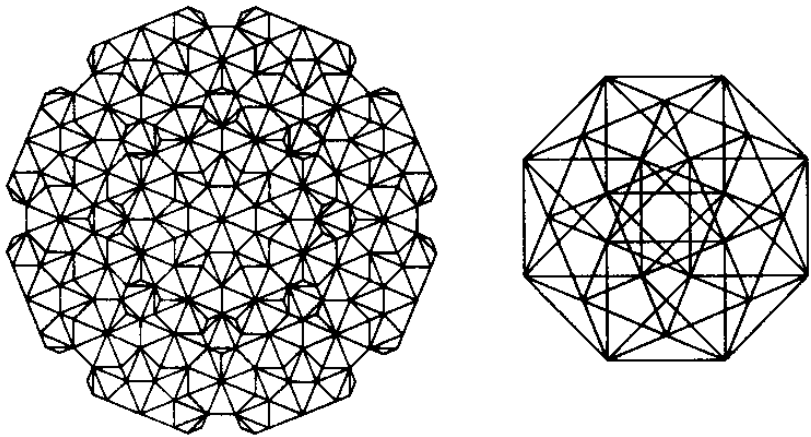}}
\caption{
 Quasiperiodic octagonal pattern (left) as obtained from the
root lattice $D_4$ and projection image of the Voronoi cell of $D_4$ in
perpendicular space.
}
\label{fig2}
\end{figure}

\begin{figure}
\centerline{ \epsfxsize=3.5in \epsfbox{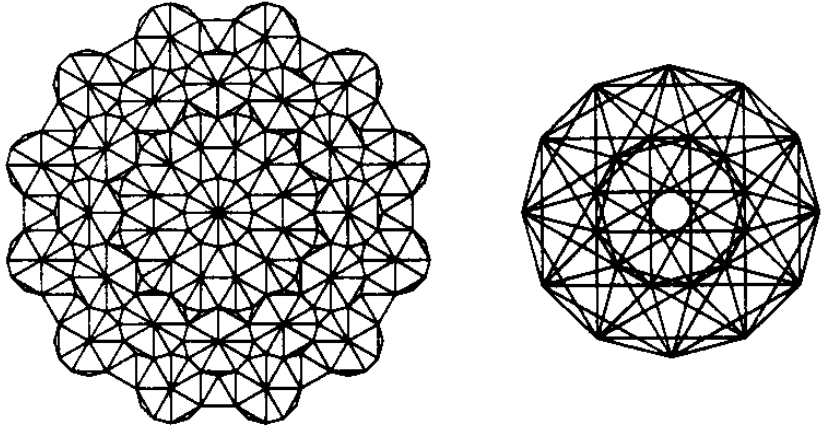}}
\caption{
Quasiperiodic dodecagonal pattern (left) as obtained from the
root lattice $D_4$ and projection image of the Voronoi cell of $D_4$ in
perpendicular space.
}
\label{fig3}
\end{figure}

The root lattices are those lattices which are generated by the root system of
a semisimple Lie algebra (cf. \cite{sloane,humpy} for details). One obtains the
list $A_n (n \geq 1)$, $\zet ^n (n \geq 2)$, $D_n (n \geq 4)$, $E_6$, $E_7$,
and $E_8$, all root lattices are constructed from by means of combinations
(i.e., of direct sums).
The holohedry of a root lattice coincides with the automorphism group of the 
corresponding root system, wherefore the symmetry is easily accessible (cf.\
\cite{humpy}). The possible angles between Voronoi vectors are multiples of
$60^0$ or $90^0$ thus weakly generalizing the hypercubic situation
(only multiples of $90^0$) which, of course, is contained.

Let us now come to the relevant examples. In 3-space, one has only the
icosahedral symmetry which is both irreducibly represented and genuinely
noncrystallographic. There are three different Fourier modules possible
with icosahedral symmetry \cite{mermin} which can, in minimal embedding, 
be obtained as a projection of the primitive (observed first \cite{sch}),
the F-type \cite{f}, and the I-type hypercubic lattice in $\rr ^6$, 
respectively. But these three lattices are just $\zet ^6$, $D_6$, 
and $D_6^R$, so we are back to root
lattices and their reciprocals (cf.\ \cite{kramerneri} and \cite{kramer}
for tiling models based on $\zet ^6$ and $D_6$, respectively).

Let us now focus on 2-D quasilattices with rotational symmetries of order
5, 8, 10, and 12, which occur in nature in form of sections through 
so-called T-phases \cite{t}, perpendicular to the symmetry axis.
The minimal embedding requires 4-D space 
(since, in each case, the Euler function is four). The root lattices provide 
these embeddings with maximal symmetry. 
The most prominent example, the Penrose pattern (Fig.\ 1a),
and its partner, the triangle pattern (Fig.\ 1b), are obtained from the root
lattice $A_4$ \cite{harg}. This covers both the fivefold and the tenfold
symmetry depending on the decoration of the tiles.

An eightfold symmetry can either be realized by means of $\zet ^4$ \cite{acht}
or by means of $D_4$ \cite{rot}, the only hypercubic centering in $\rr ^4$, 
see Fig.\ 2.
The latter has the advantage that also a pattern with twelvefold symmetry can
be obtained, see Fig.\ 3. Furthermore, one can desribe a continuous
transition \cite{rot} between the octagonal and the dodecagonal phase
within the strict dualization
scheme and Klotz construction \cite{klotz}, which all patterns shown 
are based on. Of course, the relation to different other tilings
\cite{muster} is to be investigated in detail because manifestly inequivalent
tilings (like the Penrose pattern and the triangle pattern) can exist which
nonetheless share the same Fourier module.

Presently, there is no observation of a further crystallographically forbidden
symmetry. For the observed ones, root lattices seem to provide the right
basis for structure models, and it turns out that the minimal embedding in these
cases always requires a description in a space with twice the dimension of the
quasiperiodic tiling. This phenomenon - if it is not accidental - should
have some physical reason and seems to be related  
to deflation/inflation invariance and
generalized symmetries of quasiperiodic patterns. This is to be investigated
in the future. Obviously, the tiling models based on root lattices should be
used for structure models with realistic decorations and for the calculations
of electronic and magnetic properties.
Finally, quite a lot of important and interesting questions are to be answered
on the way to a classification of quasiperiodic tiling models, some of which
might be tackled within the dualization scheme.

{\bf Acknowledgements}

This work was supported by Deutsche Forschungsgemeinschaft, Australian
Research Council, and Alfried Krupp von Bohlen und Halbach Stiftung.

\end{document}